\documentclass[showpacs,prl,twocolumn,floatfix,reprint]{revtex4}
\usepackage{graphicx,amssymb,amsmath}

\begin{document}

\title{Modeling the polycentric transition of cities}

\author{R\'emi Louf}
\email{remi.louf@cea.fr}
\affiliation{Institut de Physique Th\'{e}orique, CEA, CNRS-URA 2306, F-91191, 
Gif-sur-Yvette, France}

\author{Marc Barthelemy}
\email{marc.barthelemy@cea.fr}
\affiliation{Institut de Physique Th\'{e}orique, CEA, CNRS-URA 2306, F-91191, 
Gif-sur-Yvette, France}
\altaffiliation{CAMS (CNRS/EHESS) 190-198, avenue de France, 75244 Paris Cedex 13, France}

\begin{abstract}

Empirical evidence suggest that most urban systems experience a
transition from a monocentric to a polycentric organisation as they
grow and expand. We propose here a stochastic, out-of-equilibrium model of the
city which explains the appearance of subcenters as an effect of
traffic congestion. We show that congestion triggers the instability
of the monocentric regime, and that the number of subcenters and the
total commuting distance within a city scale sublinearly with its
population, predictions which are in agreement with data gathered for
around 9000 US cities between 1994 and 2010.

\end{abstract}


\maketitle

As cities grow, they evolve from monocentric organisations where all
the activities are concentrated in the same geographical area
--usually the central business district-- to more distributed,
polycentric organisations
~\cite{Kemper:1974,Odland:1978,Mills:1972,Griffith_PG:1981,Dokmeci:1994,McMillen:2003,Pereira:2013,Roth:2011}. Traditional
approaches in spatial economics have attempted to describe the
phenomenon within the framework of equilibrium models of the city~\cite{Fujita:1982,Fujita:book1999}. These models are traditionally
based on the concept of agglomeration economies --to explain why
economical activities tend to group-- and the spatial distribution of
wages and rents across the urban space. However, these approaches fail
at giving a satisfactory quantitative
account~\cite{Bouchaud:2008,Batty:2008} of the polycentric transition
of cities. First, they describe a city as being in an equilibrium
characterised by static spatial distributions of households and business
firms. However, the equilibrium assumption is unsupported as cities
are out-of-equilibrium systems and their
dynamics is of particular interest for practical
applications~\cite{Batty:2008}. Second, these models integrate so many
interactions and variables that it is difficult to understand the
hierarchy of processes governing the evolution of cities, which ones are fundamental and which ones are irrelevant. Yet, traffic congestion is not
explicitly taken into account in the existing models, despite being
mentioned in the economics literature as a possible reason for the
polycentric transition~\cite{McMillen:2003}. Lastly, the models do
not make any quantitative prediction and are therefore unsupported by
data.\\ 
We present in this Letter a stochastic, out-of-equilibrium model of the
city which relies on the assumption that the polycentric structure of
large cities might find its origin in congestion, irrespective of the
particular local economic details. We are able to reproduce many
stylized facts and, most importantly, to derive a general relation
between the number of activity centers of a city and its
population. Finally, we verify this relation against the employment
data from around 9000 cities in the U.S. between 1994 and 2010.


Following recent interdisciplinary efforts to construct a quantitative
description of cities and their
evolution~\cite{Makse:1995,Zanette:1997,Marsili:1998a,Marsili:1998b,Batty:book2005,Bettencourt:2007,Batty:2008},
we deliberately omit certain details and focus instead on basic
processes. We thereby aim at building a minimal model which captures
the complexity of the system and is able to account for qualitative as well as quantitive stylized
facts. The model we propose is by essence dynamical and describes the
evolution of cities' organisation as their populations increase. We
focus on car congestion --mainly due to journey-to-work commutes --
and its effect on the job location choice for individuals.

According to Fujita and Ogawa's classical model~\cite{Fujita:1982} in
spatial economics, an individual living at location $i$ will choose to
work at the location $j$ that maximises the net income after the
deduction of rent and commuting costs~\cite{Fujita:1982}
\begin{equation}
Z_0=W(j)-C_R(i)-C_T(i,j)
\end{equation} 
where $W(j)$ is the average wage paid by business firms located at $j$
(and thus varies from one location to another), $C_R(i)$ is the land
rent at $i$, and $C_T(i,j)$ is the commuting cost between $i$ and
$j$. The wage and the land rent result from the interplay between the households' and companies' locations, agglomeration effects being taken
into account. The commuting cost, on the other hand, does not usually take congestion 
into account and is taken proportionally to the Euclidean distance
$C_T(i,j) = t\, d_{ij}$ (where $t$ is the transportation cost per unit
of distance) in most studies.

The time scales involved in the evolution of cities are usually such
that the employment turnover rate is larger than the relocation
rate of households. On a short time scale, we can thus focus on the
process of job seeking alone, leaving aside the problem of the choice of
residence. In other words, we assume the coupling between both processes to be negligible: we assume that each
inhabitant newly added to the city has a random residence location and
we concentrate on understanding how such an inhabitant chooses its job
among a pool of $N_c$ potential activity centers (which we suppose are also
randomly distributed among the city). The active subcenters are then defined as the subset of potential centers which have a nonzero incoming number of
individuals. As a result of these assumptions, a worker living at $i$
will choose to work at the center $j$ such that the quantity
\begin{equation}
Z_{ij} = W(j) - C_T(i,j)
\end{equation}
is maximum. \\
We now discuss the form of the two terms $W(j)$ and
$C_T$. The problem of determining the (spatial) variations of the
average wage $W(j)$ at location $j$ is very reminiscent of some
problems encountered in fundamental physics. Indeed, the wage depends
on many different factors, ranging from the type of company, the
education level of the inhabitant, the level of agglomeration, etc.,
and in this respect is not too different from quantities that can be
measured in a large atom made of a large number of interacting
particles. In this situation, physicists found out that although it is possible
to write down the corresponding equations, not only is it
impossible to solve them, but it is also not really useful. In fact, they
found out~\cite{Dyson:1962} that a statistical description of these
systems, relying on random matrices could lead to predictions that agree with experimental results. We wish to import in spatial economics
this idea of replacing a complex quantity such as wages, which depends on
so many factors and interactions, by a random one. We therefore decide to
account for the interaction between activity centers and people by
taking the wage as proportional to a random variable $\eta_j \in \left[ 0,1\right]$
such that $W(j) = s\, \eta_j$ where $s$ defines the maximum attainable
average wage in the considered city.

As for the transportation cost $C_T(i,j)$, we choose it to be
proportional to the commuting time between $i$ and $j$. In a typical
situation where passenger transportation is dominated by personal
vehicles, this commuting time not only depends on the distance between
the two places but also on the traffic between $i$ and $j$, the vehicle capacity of
the underlying network, and its resilience to congestion. The Bureau of
Public Road formula~\cite{Branston:1976} proposes a simple form taking
all these factors into account. In our framework, it leads to the following expression for the 
commuting costs
\begin{equation}
C_T(i,j) =  t\, d_{ij} \left[ 1 + \left( \frac{T_{ij}}{c} \right)^{\mu} \right]
\end{equation}
where $T_{ij}$ is the traffic per unit of time between $i$ and $j$ and $c$
is the typical capacity of a road (taken constant here). The quantity
$\mu$ is a parameter quantifying the resilience of the transportation
network to congestion. We further simplify the problem by assuming
that the traffic $T_{ij}$ is only a function of the subcenter $j$ and
therefore write $T_{ij}=T(j)$ the total traffic incoming in the subcenter
$j$ (see Supplementary Material~\cite{SM} for a short discussion).

In summary, our model is defined as follows. At each time step, we add
a new individual $i$ located at random in the city, who will
choose to work in the activity area $j$ (among $N_c$ possibilities
located at random) such that the following quantity
\begin{equation}
Z_{ij} = \eta_j - \frac{d_{ij}}{\ell} \left[ 1 + \left( \frac{T(j)}{c} \right)^{\mu} \right]
\label{eq:cost_function}
\end{equation}
is maximum (we omitted irrelevant multiplicative factors). The quantity $\ell = s/t$ is interpreted as the maximum effective
commuting distance that people can financially withstand.

\begin{figure}
\includegraphics[width=0.4\textwidth]{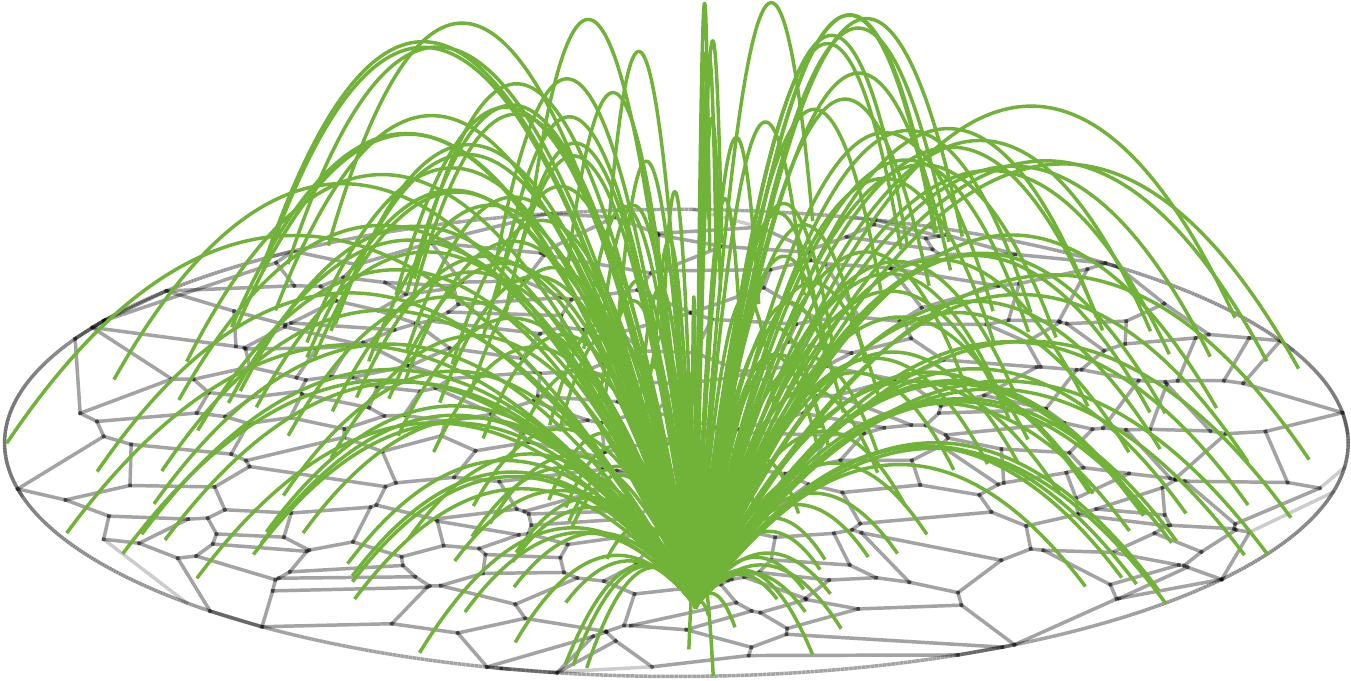}
\includegraphics[width=0.4\textwidth]{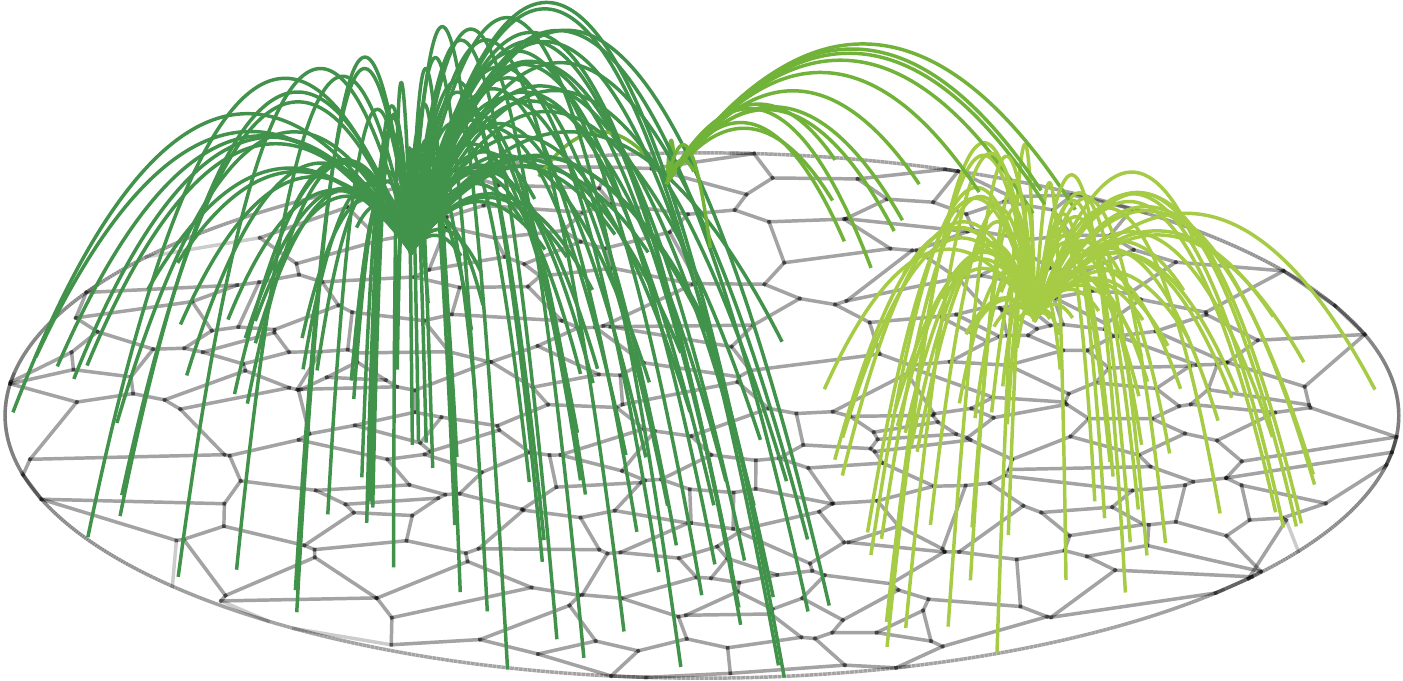}
\includegraphics[width=0.4\textwidth]{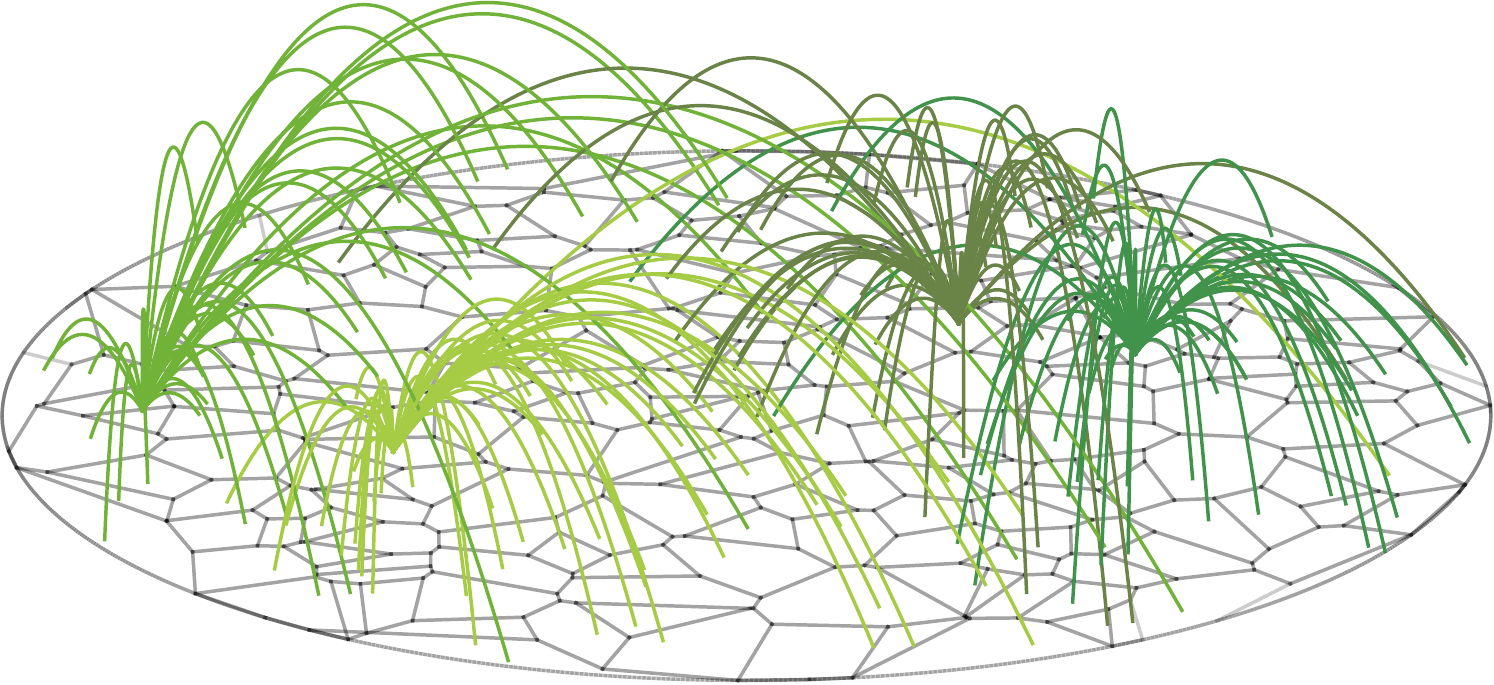}
\caption{The monocentric (top), distance-driven polycentric (middle),
  and attractivity-driven polycentric (bottom) regimes as produced by
  our model. Each link represents a commute to an activity center. \label{fig:model_results}}
\end{figure}

Depending on the relative importance of wages, distance and congestion, the
model predicts the existence of three different regimes: the monocentric
regime (top Fig.~\ref{fig:model_results}), the distance-driven polycentric (middle Fig.~\ref{fig:model_results}) regime
and the attractivity-driven polycentric (bottom Fig.~\ref{fig:model_results}) regime.

From now on, we will assume that $\ell$ is large enough so that a
monocentric state exists for small values of the population. In this
regime, the value of $\eta$ prevails and the monocentric state evolves
to an attractivity-driven polycentric structure as the population
increases (if $\ell$ is too small, the monocentric regime does not
exist --see the Supplementary material~\cite{SM} for more details on
these points).  Starting from a small city with a monocentric
organisation, the traffic is negligible and $Z_{ij}\approx \eta_j$
which implies that all individuals are going to choose the most
attractive center (with the largest value of $\eta_j$, say
$\eta_1$). When the number $P$ of households increases, the traffic
will also increase and some initially less attractive centers (with
smaller values of $\eta$) might become more attractive, leading to the
appearance of new subcenters characterized by a nonzero number of
commuters. More precisely, a new subcenter $j$ will appear when for an
individual $i$, we have $Z_{ij}>Z_{i1}$. The traffic so far is
$T(1)=P$ and $T(j)=0$ which leads to the equation
\begin{equation}
\eta_j-\frac{d_{ij}}{\ell}>\eta_1-\frac{d_{i1}}{\ell}\left[1+\left(\frac{P}{c}\right)^\mu\right]
\end{equation}
We assume that there are no spatial
correlations in the subcenter distribution, so that we can make the
approximation $d_{ij}\sim d_{i1}\sim L$. The new subcenter will thus
be such that $\eta_1-\eta_j$ is minimum implying that it will have the
second largest value denoted by $\eta_j=\eta_2$. For a uniform distribution
(details of the calculation can be found in the Supplementary
Material~\cite{SM}, section 2), on average
$\overline{\eta_1-\eta_2}\simeq 1/N_c$ leading to a critical value for
the population
\begin{equation}
P^*= c \left( \frac{\ell}{L N_c} \right)^{1/\mu}
\end{equation}
Whatever the system considered, there will therefore \emph{always} be
a critical value of the population above which the city becomes
polycentric (which can be smaller than one, in which case there is no
monocentric regime at all, see the Supplementary Material~\cite{SM}). The monocentric regime is therefore fundamentally
unstable with regards to population increase, which is in agreement with the fact that no major city in
the world exhibits a monocentric structure. We note that the smaller
the value of $\mu$ (or the larger the value of the capacity $c$), the
larger the critical population value $P^*$ which means that cities
with good road systems are capable of absorbing large traffic show a
monocentric structure for a longer period of time.


Having established that cities will eventually adopt a polycentric
structure, we can wonder how the number of subcenters varies with the
population. We compute the value of the population at which 
the $k^{th}$ center appears. We still assume that we are in the attractivity-driven regime and that, so far, $k-1$
centers have emerged with $\eta_{1} \geq \eta_{2} \geq ... \geq
\eta_{k-1}$~\cite{SM}, with a number of commuters $T(1), T(2), ...,
T(k-1)$, respectively. The next worker $i$ will choose the center $k$ if
\begin{equation}
Z_{ik} > \max_{j \in \left[1,k-1\right]} Z_{ij}
\end{equation}
which reads
\begin{equation}
\eta_k - \frac{d_{ik}}{\ell} > \max_{j \in \left[1,k-1\right]} \left\{
\eta_j - \frac{d_{ij}}{\ell} \left[ 1 + \left(
  \frac{T(j)}{c}\right)^\mu\right] \right\}
\end{equation}
The distribution of traffic $T(j)$ is narrow~\cite{SM}, which means that all the centers have roughly the same number of
commuters $T(j) \sim P/(k-1)$. As above we also assume that the
distance between the workers' households and the activity centers is
typically $d_{ij} \sim d_{ik} \sim L$. The previous expression now
reads
\begin{equation}
\frac{L}{\ell} \left( \frac{P}{(k-1)\,c} \right)^{\mu} > \max_{j \in
  \left[1,k-1\right]} \left( \eta_j \right) - \eta_k
\end{equation}
Following our definitions, $\max_{j \in \left[1,k-1\right]} \left(
\eta_j \right) = \eta_1$. According to order statistics, if the
$\eta_j$ are uniformly distributed, we have on average
$\overline{\eta_1 - \eta_k} = (k-1)/(N_c+1)$. It follows from these
assumptions that (1) the $k^{th}$ center to appear is the $k^{th}$ most attractive one (2) the average value of the population $\overline{P}_k$ at
which the $k^{th}$ center appears is given by:
\begin{equation}
\overline{P}_k = P^* \left( k-1 \right)^{\frac{\mu+1}{\mu}}
\label{eq:prediction}
\end{equation}
Conversely, the number $k$ of subcenters scales sublinearly with population as
\begin{equation}
k \sim \left( \frac{\overline{P}}{P^*} \right)^{\frac{\mu}{\mu + 1}}
\end{equation}
It is interesting to note that this result is robust: the dependence
is sublinear, \emph{whatever the distribution} of the random variable
$\eta$ (see the Supplementary Material~\cite{SM} for a discussion on this
point). We can therefore conclude that, probably very generally and
under mild assumptions, the number of activity subcenters in urban
areas scales sublinearly with their population where the prefactor and
the exponent depend on the properties of the transportation network of
the city under consideration. 

A previous study~\cite{Samaniego:2008} showed that the total
miles driven daily in a city --- the `total commuting
distance'---scales with the population as $L_{tot} \sim P^\gamma$
where $\gamma \in \left] 0.5, 1\right[$, which the authors interpreted
as cities having neither totally centralized nor totally
decentralized structures. We can discuss this result within the
framework of our model in the following way. If the system was in the
pure attractivity-driven regime, we would have $L_{tot} \sim P$. But, if we
assume that we are in an intermediate regime where
Eq.~\ref{eq:prediction} holds, and where the system exhibits spatial
coherence~\cite{SM}, we can write the total length of the commutes as
\begin{equation}
L_{tot} \sim P \frac{L}{\sqrt{k}}
\end{equation} 
Inverting the result from Eq.~\ref{eq:prediction} we therefore get
\begin{equation}
L_{tot} \sim P^{1-\frac{\beta}{2}}
\label{eq:beta}
\end{equation}
where $\beta = \frac{\mu}{\mu+1} \in \left[0 , 1\right]$. Our model is
thus consistent with the fact that the total traveled miles scales with
population with a non-trivial exponent comprised in $[0.5,1]$.\\


\begin{figure*}
\includegraphics[width=\textwidth]{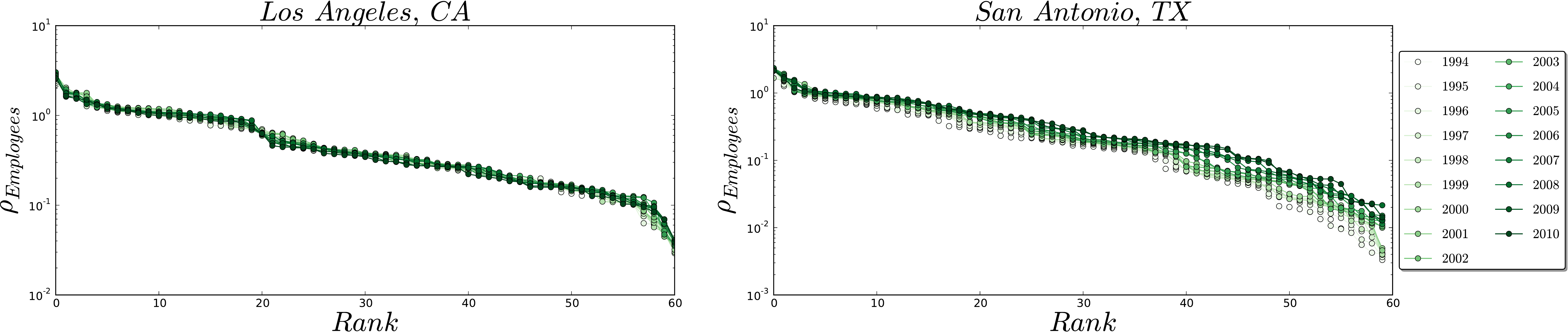}
\caption{Rank-plot for the employment density (in employees per
  $km^2$) in Los Angeles, CA (left) and San Antonio, TX (right)
  between 1994 and 2010. See the Supplementary Material~\cite{SM} for more details. \label{fig:rank_plots}}
\end{figure*}

\begin{figure}
\includegraphics[width=0.5\textwidth]{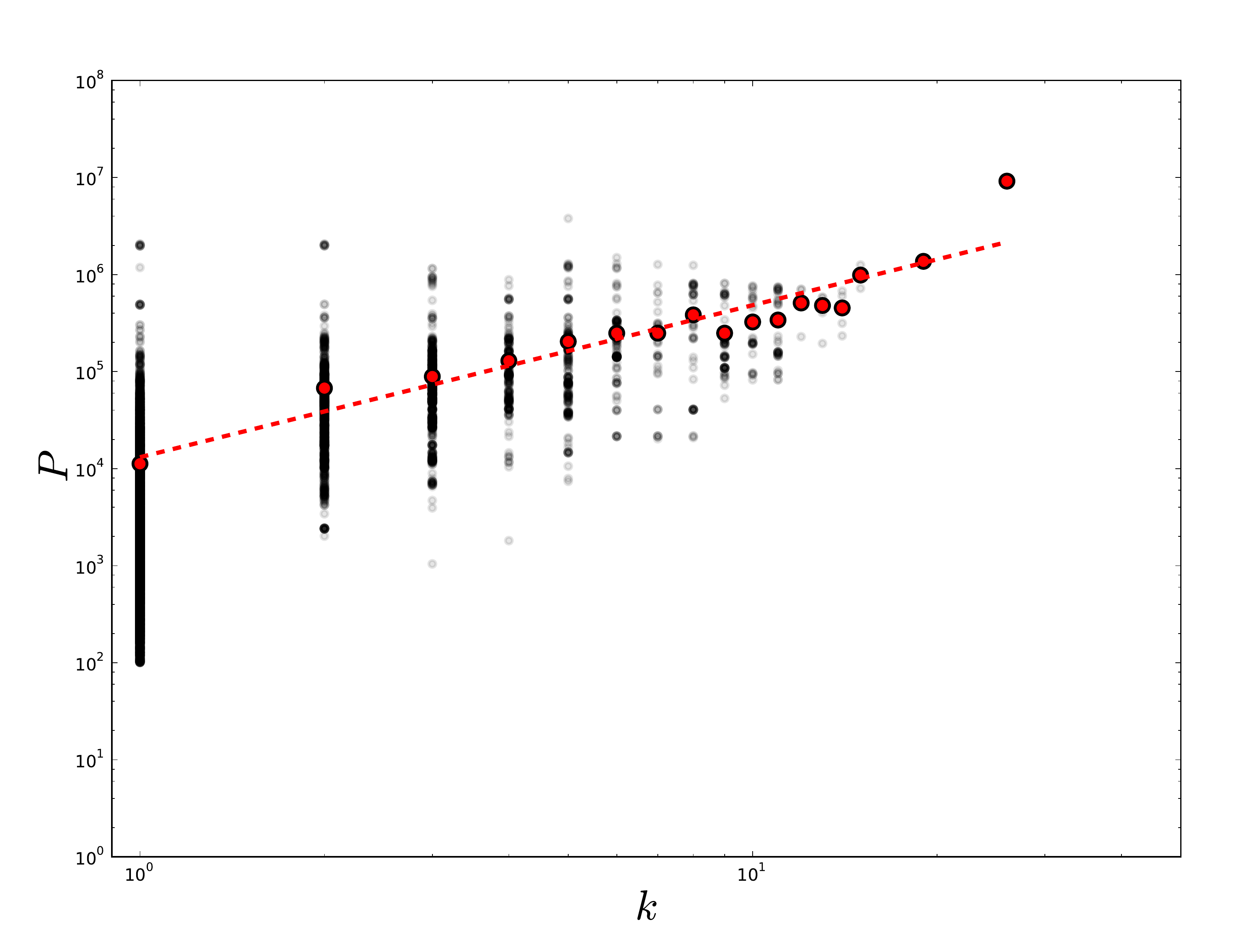}
\caption{Scatter plot of the estimated number of subcenters versus the
  population for about 9000 cities with population over 100 people in
  the US. The red dots represent the average population for a given
  number of subcenters. We fit this average with a power-law
  dependence (represented by a red dashed line) giving an exponent
  $\delta=1.56\pm 0.15$ ($r^2=0.87$). See the Supplementary Material~\cite{SM} for more details on the computation of $k$ and the robustness of the results. \label{fig:data}}
\end{figure}

We now test the prediction given by Eq. (\ref{eq:prediction}). For
that purpose, we collected data for the number of employees per zip
code in the United States that were collected over a span of 16 years, between 1994 and
2010~\cite{ZBPdata}, as well as the population of all cities in the US
between 1994 and 2010~\cite{CensusData}. We estimate the number of
subcenters by constructing the rank plot of the employment density
$\rho$ (number of employees per $km^2$) for each Zip Code of a given
urban area~\cite{Griffith:1981,Dokmeci:1994}. These plots display a decay as fast as an exponential
(Fig.~\ref{fig:rank_plots}) which implies that there exists a natural
scale for the rank, that we interpret here as the typical number of
activity centers. It also implies that any reasonable method should
give an estimate of the number of subcenters of the same order of
magnitude (which would not be the case for slowly decaying functions
such as power laws for example). We first note that for some cities
--typically large ones with stable populations
(Fig.~\ref{fig:rank_plots}, left)-- the employment spatial statistics
remained stable over the period of study. For other cities, we observe
large variation of the number of subcenters
(Fig.~\ref{fig:rank_plots}, right). We then plot (Fig.~\ref{fig:data})
the population $P$ of cities (with population $P>100$) versus the
estimated number of subcenters $k$ (the dispersion in the scatter plot
probably results from the fact that different cities have different
resilience levels to congestion). On average, we observe a power law
dependence with exponent $\delta=1.56 \pm 0.15$ (the result is robust
with regards to the estimate of $k$, see the Supplementary
Material~\cite{SM} for more details). Inverting this relation gives us
the number of subcenters as a function of the population
\begin{equation}
k \sim P^{\beta}
\end{equation}
with $\beta \sim 0.64$. This result is strikingly in agreement with
the prediction given by our model: the number of subcenters in a city
scales sublinearly with its population.\\

Using the measured value of $\beta$ and Eq. (\ref{eq:beta}) we can estimate the exponent of
the scaling of $L_{tot}$ with the population and find $L_{tot} \sim
P^{\,0.68}$ which agrees very well with the value $0.66$ measured
in~\cite{Samaniego:2008} directly on the data of the daily total miles
driven in more than $400$ cities in the US.\\

While agglomeration economies seem to be the basic process explaining
the existence of cities and their spectacular resilience, this study
brings evidence that congestion is the driving force that tears them
apart. The nontrivial spatial patterns observed in large cities can
thus be understood as a result of the interplay between these
competing processes. We believe that the present model represents an
important step towards a quantitative, predictive science of
cities. More generally, this microscopic approach is an interesting
example of an out-of-equilibrium model: it is governed by local
optimization with saturation effects, leads to different regimes, and
is characterized by nontrivial dynamical exponents. In this respect,
we believe that this discrete approach might be of use in the study of
pattern formation in biology --which has been so far explored from a
global optimisation perspective~\cite{Ashton:2005}-- or used as a coarse-grained approach to reaction-diffusion equations with a density-dependent diffusion coefficient~\cite{Cates:2012} to compute quantities that
are out of reach within the current methods.

Acknowledgements-- MB thanks H. Berestycki, A. Flammini, and K. Mallick
for stimulating discussions. MB acknowledges funding from the EU Commission  through project
EUNOIA (Project No. FP7-DG.Connect-318367).

\section{Supplementary Materials}

\subsection{Details on the simulations}

The simulation results presented in these supplementary materials are
obtained in the following way. We first distribute randomly a number
$N_c$ of potential activity centers uniformly in the unit
circle. Then, at each time step, we add a worker $i$ at a random
position in the circle, and compute the cost functions $Z_{ij}$ for
all potential activity centers $j$. We then connect the worker with
the center maximizing the cost function, and add one to the value of the traffic
corresponding to this center. We repeat this procedure until all
workers are connected to a center.\\

Strictly speaking, traffic congestion do not arise only at the exact location of the activity center, but in an area of typical size $R$ around that center. People who are not commuting to this specific center, but who have to travel close to it when commuting to their working place, might add to the traffic in the surroundings of this center. Therefore, we should consider the traffic $T_R(j)$ in the area around the center $j$, that is to say the number of people whose commute path crosses the area of size $R$ around this center. For the sake of simplicity and clarity, we choose to ignore this in our model and leave it for further investigation.

\subsection{Attractivity-driven and distance-driven polycentric regimes}
\subsubsection{Definition and consequences}
Fig. 1 in the main text suggests the existence of two different types of polycentric regimes that we called attractivity-driven and distance-driven polycentric regimes. In the attractivity-driven regime, individuals decide to work at the most attractive center, provided that the traffic is not too large. Therefore, as the traffic increases, new centers are going to appear in the decreasing order of attractivity. In the distance-driven regime, however, individuals tend to connect to the closest center, and thus the centers should appear in a random order.\\

More specifically, the attractivity-driven regime appears when the term $\frac{d_{ij}}{\ell}\left[1+\left(\frac{T(j)}{c}\right)^\mu\right]$ is negligible compared to the attractivity $\eta_j$ at small values of traffic. Then, maximising the value of $Z_{ij}$ amounts to maximising the attractivity of the center $j$. This will typically be the case when:

\begin{equation}
\ell \gg \ell^* = L
\end{equation} 

On the other hand, when $\ell \ll \ell^*$, the attractivity of the centers becomes irrelevant and maximising $Z_{ij}$ amounts to connecting the closest center $j$ to $i$. An important consequence is that the assumptions used to derive Eq. 6 and Eq. 10 in the main text are not valid in the distance-driven regime. In fact, there cannot be any stable monocentric state in the distance-driven regime and we start from the beginning (after a few iterations) with a polycentric state. Therefore, the difference between attractivity-driven and distance-driven polycentric regime draws the distinction between a system which grows from a single center, and a system where several centers appear from the beginning. This may be interpreted as the emergence of a single city when people can afford to travel a longer distance than the extension of the system (attractivity-driven), and of a system of cities when they cannot (distance-driven regime).\\

In between these two regime lies an intermediate regime, which is neither completely driven by space, not completely driven by the attractivity. This interplay between space and attractivity makes this intermediate regime difficult to explore analytically, we thus limit ourselves in this Letter to the two extreme regimes.

\subsubsection{The transition is towards the attractivity-driven regime}

We now show that if the parameters are such that there exists a monocentric state, then the system necessary evolves to an attractivity-driven polycentric structure. Let us assume that the system is in a monocentric state: $P$ individuals commute to the center $j=1$ characterised by the attractivity $\eta_1$. Let us assume that the monocentric state is unstable and that the $P+1$th individual $i$ has two different possibilities. First, she could choose to go to the next most attractive subcenter characterised by the attractivity $\eta_2$, the largest attractivity among all the remaining potential centers. This translates into
\begin{equation}
Z_{i2}=\eta_2-\frac{d_{i2}}{\ell}>Z_{i1}=\eta_1-\frac{d_{i1}}{\ell}\left[1+\left(\frac{P}{c}\right)^\mu\right] 
\end{equation}  
In general, the most attractive center is at a random fraction of the system size $L$, that is to say $d_{i1} \sim d_{i2}\sim L$. We are then led to Eq. 6 of the main text (see further in the Supplementary Material for a derivation)
\begin{equation}
P^*=c\left(\frac{\ell}{LN_c}\right)^{1/\mu} 
\end{equation}
The second possibility is for the individual to choose the closest center, characterized by $\eta_j$. We then 
have
\begin{equation}
Z_{ij}=\eta_j-\frac{d_{ij}}{\ell}>Z_{i1}=\eta_1-\frac{d_{i1}}{\ell}\left[1+\left(\frac{P}{c}\right)^\mu\right]
\end{equation}
Since $j$ is the closest subcenter to $i$ we can write $d_{ij}\sim L/\sqrt{N_c}$ and after a simple calculation we obtain a new value for $P^*$ (in the limit of large $N_c$)
\begin{equation}
P'^*=c\left(\frac{\ell}{2L}\right)^{1/\mu}
\end{equation}

We immediately observe here that for all values of the parameters, we always have $P^*<P'^*$. This result indicates that if a monocentric state
exists, the transition is always towards an attractivity-driven polycentric structure. Or conversely, that there cannot be any stable monocentric state in the distance-driven regime.

\subsubsection{Numerical verifications}

\paragraph{Ordering in the apparition of centers}
In order to verify the ordering in the apparition of secondary centers --and therefore the existence of two distinct regimes--, we increase the population in a certain configuration until all the the centers are populated. Each time a new center is populated, we note its attractivity. At the end, we compare the list obtained with the list of all the attractivities of the centers in decreasing order and compute Kendall's $tau$ function defined by:

\begin{equation}
\label{kendalltau}
\tau = \frac{N_c - N_d}{\frac{n(n-1)}{2}}
\end{equation}

where $n$ is the number of centers, $N_c$ the number of concordant pairs and $N_d$ the number of discordant pairs. A pair $i,j$ is said to be concordant if $x_i>x_j$ and $y_i >y_j$ or if $x_i < x_j$ and $y_i < y_j$. It is said to be discordant otherwise. One can see that if the orderings are identical we have $\tau=1$, if they are completely opposite we have $\tau=-1$ and $\tau=0$ if there is no correlation between the two lists.\\

\begin{figure}[!h]
\centering
\includegraphics[width=0.45\textwidth]{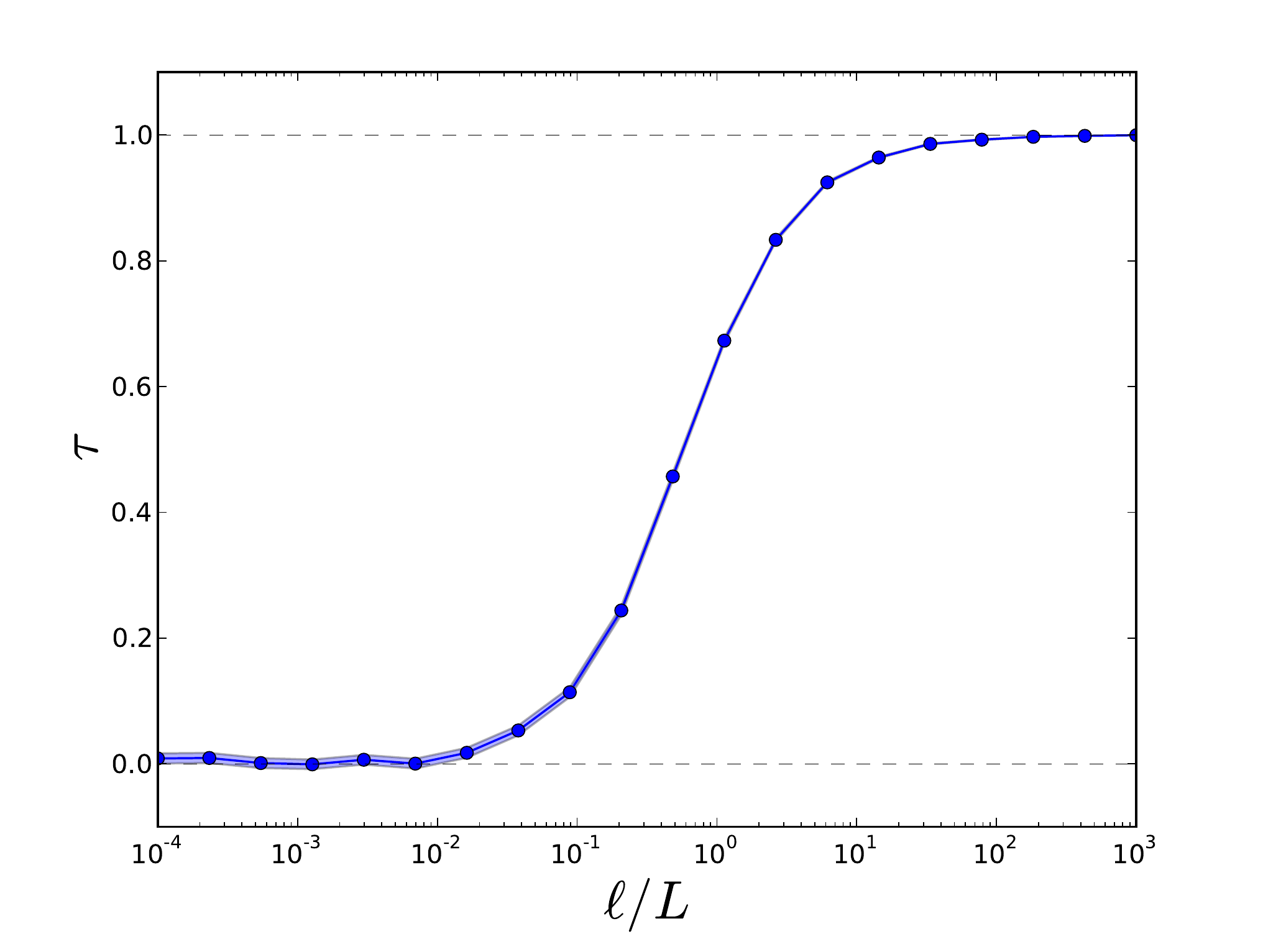}
\caption{Evolution of Kendall's $\tau$ with $\ell / L$ for $N_c=10$, $\mu=4$, and $c=1$. The curve shows the average for 1000 configurations, and the standard deviation is shown by the shaded area. The dotted black lines highlight the extreme values $\tau=0$ and $\tau=1.$~\label{fig:kendall}}
\end{figure}

Fig.~\ref{fig:kendall} shows the evolution of $\tau$ with the value of $\ell$. We see that we have $\tau=0$ for small values of $\ell$, indicating that the centers appear in a random order, and that we are in a distance-driven regime. On the other hand, for large values of $\ell$ we have $\tau=1$, indicating that the centers appear in increasing order of their attractivity, thus that we are in an attractivity-driven regime. The study of the nature of the transition between the two regimes goes beyond the scope of this paper, and we leave it for further investigation.\\

\begin{figure}[!h]
\centering
\includegraphics[width=0.45\textwidth]{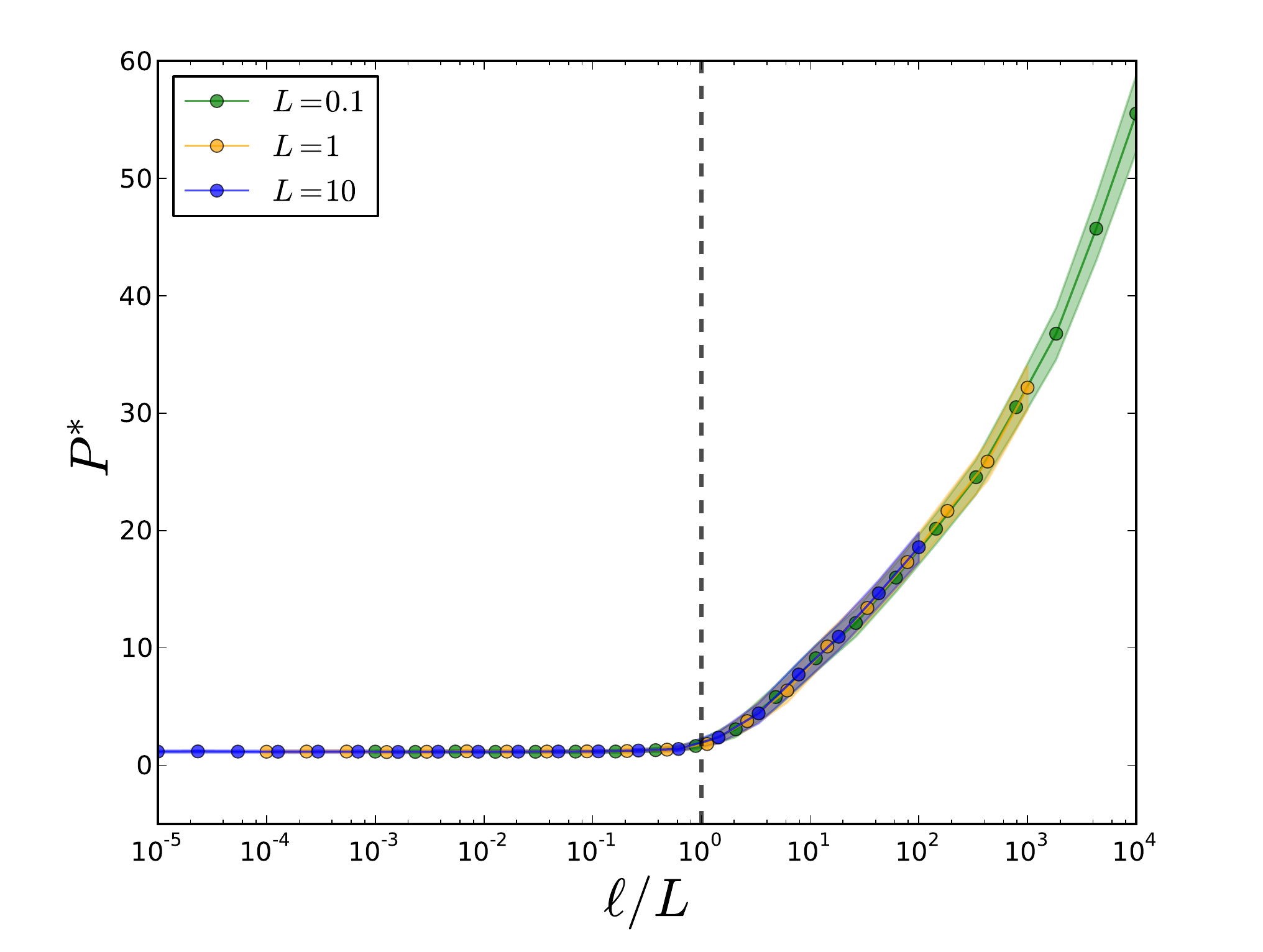}
\caption{Evolution of $P^*$, the value of population at which a second center appears, with $\ell/L$ for several values of $L$ ($\mu=4$,$N_c=10$,$c=1$, average of 10000 configurations). All the plots collapse on the same curve, and indicate that there is no stable monocentric state for $\ell / L < 1$.~\label{fig:Pstar}}
\end{figure}

\paragraph{Existence of a monocentric state}
We confirm numerically the absence of a stable monocentric state in the distance-driven regime, and its existence in the attractivity-driven regime. We plot on Fig.~\ref{fig:Pstar} the value of $P^*$, the population at which a second center appears for different values of $\ell/L$. We see that when $\ell < L$, $P^* \approx 1$, which means that there is no stable monocentric state. On the other hand, when $\ell > L$, $P^*$ is different from $1$, there exists a stable monocentric regime. We show later in these Supplementary Materials that we can compute an analytical expression for $P^*$ when $\ell \gg L$.

\subsection{Spatial coherence}

In the derivation of Eq. 13 in the main text, we assume that cities are systems such that $\ell > L$ (and Eq. 11 holds) while still exhibiting some spatial coherence. By spatial coherence, we mean that even if we are in a regime both controlled by the attractivity and space. In particular, it implies that the overlap between the basins of attraction of each subcenter is relatively small. In order to justify this assumption, we study numerically the behaviour of the average commute distance $\overline{d}$.\\

\begin{figure}
\centering
\includegraphics[width=0.45\textwidth]{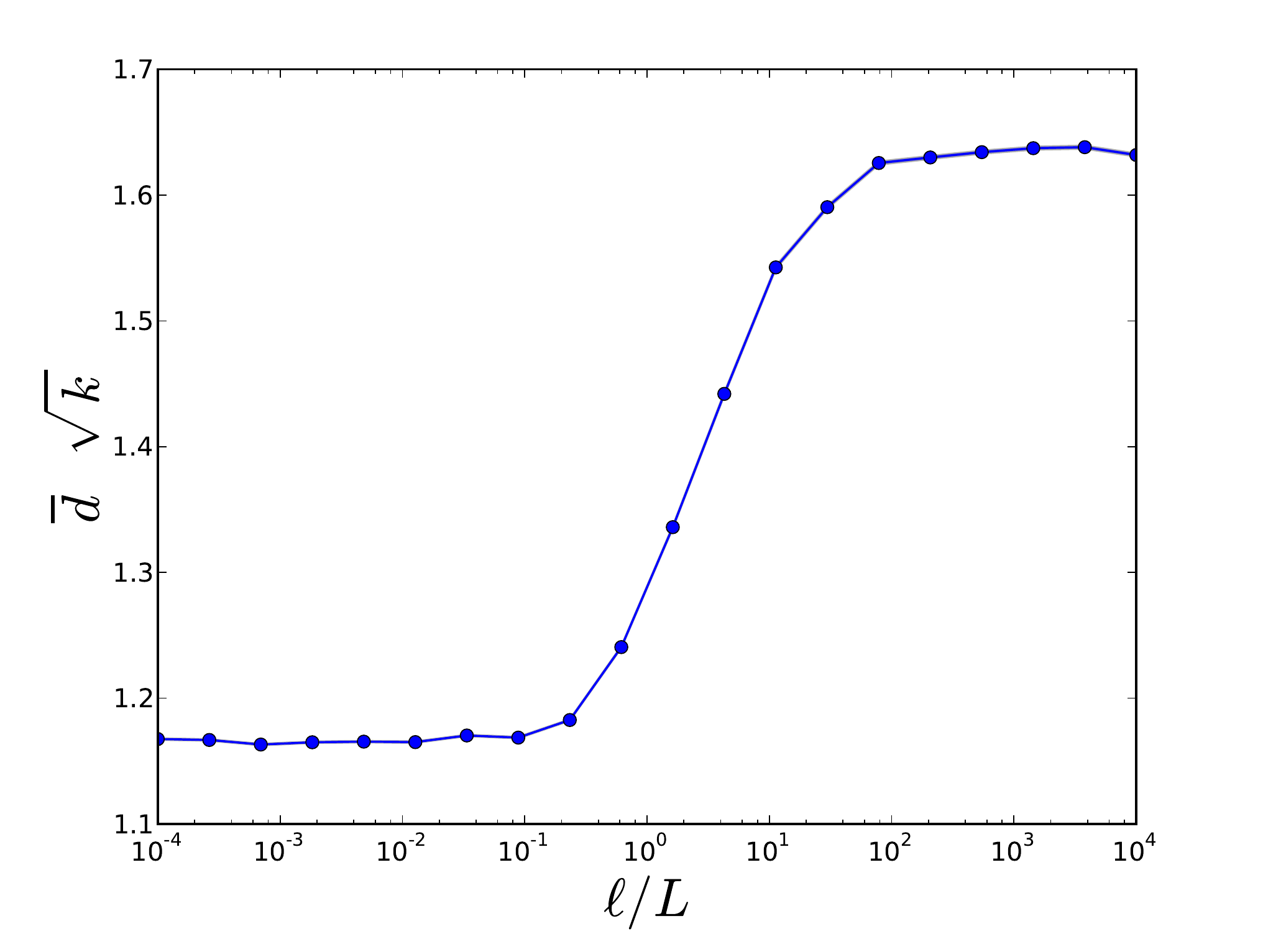}
\caption{Evolution of $\overline{d}\: \sqrt{k}$ with the ratio $\ell /L$ computed over 10000 generated systems ($\mu=4$, $c=100$, $N_c=10$). We generate the systems such that all the centers are occupied, $k=N_c$.
For small values of $\ell / L$ we are in the distance-driven regime, people commute to the closest center and the ratio is close to one. On the other hand, when $\ell /L$ is large, people commute to the most attractive center disregarding of the distance and the ratio is close to $\sqrt{N_c} L/2$. In the intermediate regime, the transition is relatively slow, meaning the system keeps a spatial coherence for a reasonable range of $\ell / L$.\label{fig:coherence_length}}
\end{figure}

If we are in a regime where the individuals connect to the closest center, we have $\overline{d} \sim L / \sqrt{k}$ where $k$ is the number of active centers. On the other hand, if we are in a regime where individuals tend to connect to the most attractive centers, then the average commute distance scales as $\overline{d} \sim L$.
We thus plot $\sqrt{k}\;\overline{d}$ as a function of $\ell/L$ on Fig.~\ref{fig:coherence_length}. We can see that $\sqrt{k}\;\overline{d}$ increases as $\ell/L$ increases: people travel smaller distances in the distance-driven regime that in the attractivity-driven regime. Yet, for values of $\ell/L$ which are not too large (typically 10) the value of the average commute distance lies between the two extremes, closer to the value corresponding to the distance-driven regime, which supports the assumption that has been made in the main text to derive Eq. 13: up to large values of $\ell/L$ for which we know from figure 1 that we are in the attractivity driven regime, we can still use the assumption of the existence of well-defined basins of attraction for each subcenter with small overlap with each other.

\subsection{Computation of $P^*$ in the attractivity-driven regime}

\subsubsection{Analytical derivation}

In this section, we explicit the details of the calculation of the population $P^*$ for which a first secondary center appears in a city within our model. We are here interested in the regime $\ell \gg L$ in which the monocentric state is stable for small values of the population (cf Fig.2). We assume that the system is in a situation where all the existing workers are connected to a single center $1$, the most attractive of all the possible centers. We would like to know what happens when we add the next workers, that is to say whether a secondary center is going to appear, and if so, for what value of the population. \\
A secondary center $k$ will appear when the $(P+1)^{th}$ worker $i$ is
added to the system if:

\begin{equation}
Z_{ik} \geq Z_{i1}
\end{equation}

Following our assumptions, the traffic to $1$ is given by $T(1) = P$ and the traffic to $k$ by $T(k)=0$. Therefore, the previous equation, in combination with the definition of $Z_{ij}$ reads:

\begin{equation}
\eta_k - \frac{d_{ik}}{\ell} \geq  \eta_1 - \frac{d_{i1}}{\ell}\left[ 1 + \left( \frac{P}{c} \right)^\mu \right]
\end{equation}

Which can be written as:

\begin{equation}
\left( \frac{P}{c} \right)^\mu \geq \frac{\ell}{d_{i1}} \left[ \eta_1-\eta_k + \frac{d_{ik}}{\ell}\right] - 1
\end{equation}

Which is so far an exact result. In order to derive the average value of $P^*$, we make some further assumption. First, because households and activity centers are distributed at random, the distance to the different centers is taken to be approximately $d_{i1} \sim d_{ik} \sim L$, the typical size of the system. Also, we know from order statistics~\cite{David:1970} that in the case of a random variable $\eta$ uniformly distributed in $\left[ 0,1 \right]$, such that $\eta_1 \geq \eta_2 \geq \dots \geq \eta_k$ we have 

$$\overline{\eta_1-\eta_k} = \frac{k-1}{N_c+1}$$

on average. Therefore, the first secondary center to appear is necessarily the center $k$ which maximises $\overline{\eta_1-\eta_k}$, that is to say the second most attractive center with attractivity $\eta_2$. When $N_c \gg 1$ we find that a secondary center appears when the population $P$ of the city is such that:

\begin{equation}
P \geq P^* = c \left( \frac{\ell}{L N_c} \right)^{1/\mu}
\label{eq:pstar}
\end{equation}

Thus, if the parameters are such that a monocentric configuration can exist in the first place, there will always be a value of the population over which a secondary center is going to appear.\\

\subsubsection{Numerical verification}
\begin{figure}
\centering
\includegraphics[width=0.45\textwidth]{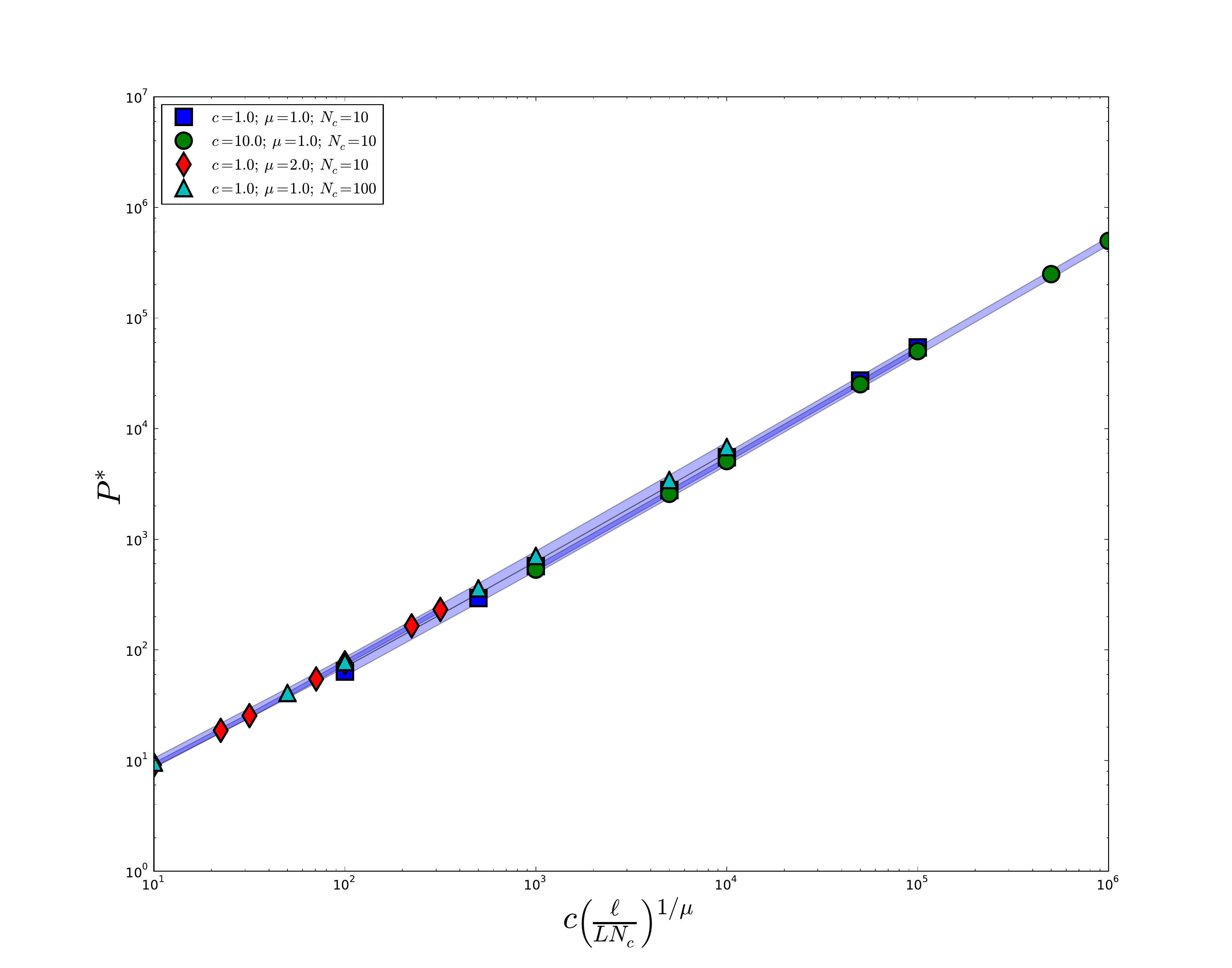}
\caption{Verification of the expression for $P^*$. Shows the average of the measured $P^*$ on many simulated configurations of the system as a function of $c \left( \frac{\ell}{L N_c} \right)^{1/\mu}$, for different values of the parameters. The points all collapse on the same curve, confirming our expression for $P^*$.\label{fig:collapse_Nstar}}
\end{figure}

In order to check the validity of our assumptions and thus of the previous formula, we numerically generate a lot of systems for different values of $\ell$ , $L$, $N_c$ and $\mu$ and determine the population $P^*$ at which a secondary center appears. We then plot the average of $P^*$ as a function of $c \left( \frac{\ell}{L N_c} \right)^{1/\mu}$ (Fig.~\ref{fig:collapse_Nstar}) and see that all the plots collapse on the same curve, confirming the above expression of $P^*$.

\subsubsection{Existence of the monocentric regime}

We saw previously that there always exists a value of the population size over which a secondary center appears. The same expression for $P^*$ also implies that even in the regime $\ell \gg L$ there is a range of parameters for which the monocentric regime simply does not exist, that is when $P^* < 1$. Assuming $N_c$ and $\mu$ are fixed, for every value of the capacity $c$ there exists a value $\tilde{\ell}$  of $\ell$ given by :

\begin{equation}
\tilde{\ell} = L\;N_c\;\left(\frac{1}{c}\right)^\mu
\end{equation}

under which there is no monocentric regime and the system exhibits a polycentric structure from the beginning of its evolution. Therefore, we witness a transition from a monocentric to a polycentric regime only if the parameters are such that $\ell > \ell*$ and $\ell > \tilde{\ell}$

\subsection{Some details about $P_k$}

\subsubsection{Distribution of $T(j)$}

\begin{figure}[!h]
\centering
\includegraphics[width=0.45\textwidth]{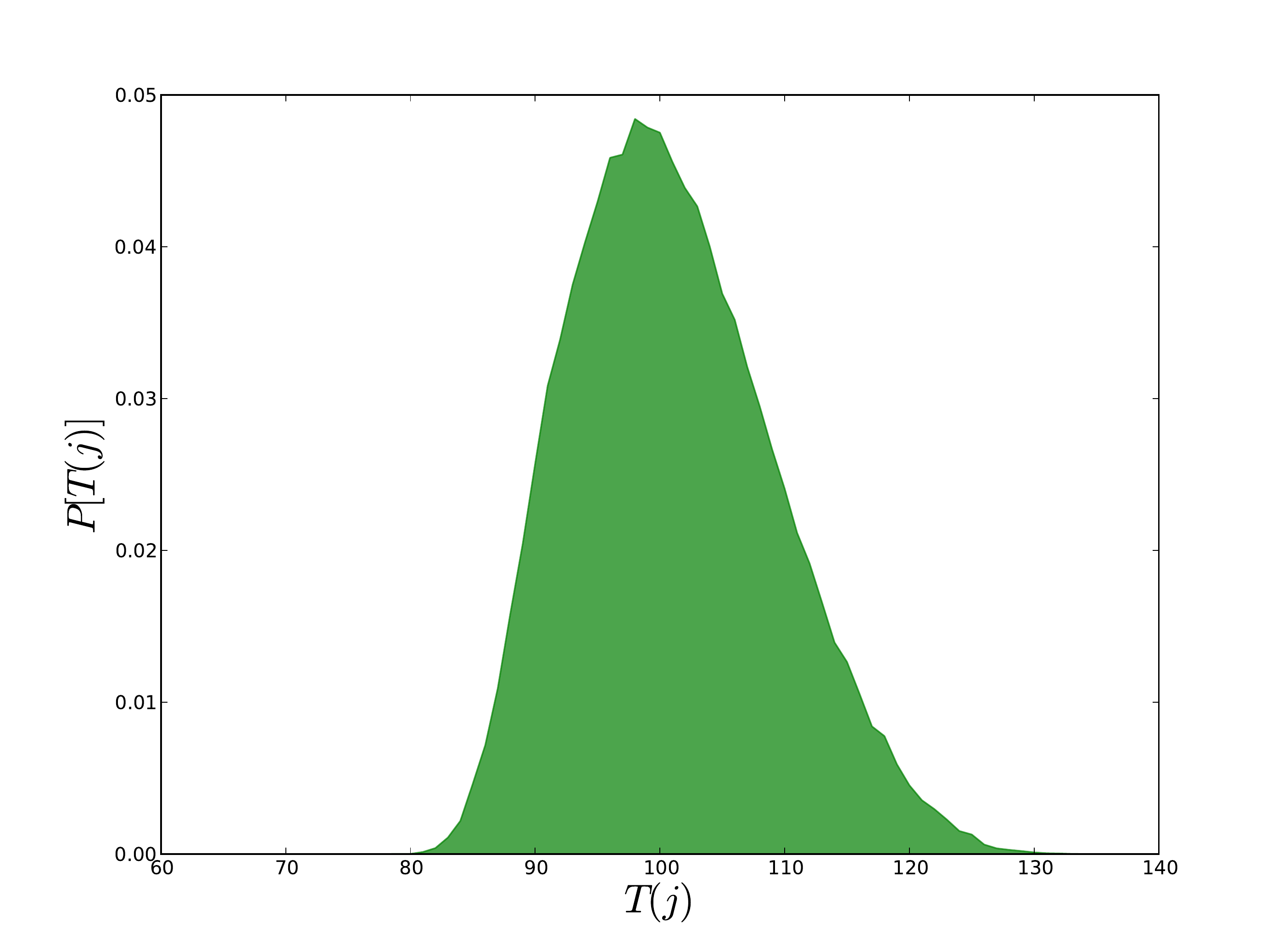}
\caption{Distribution of $T(j)$ obtained from the simulation of 10000 systems with the parameter $\mu=4$, $c=1$, $L=1$, $\ell=10^4$, $N_c=10$ and $P=10^3$\label{fig:traffic_distribution}}
\end{figure}

In the derivation of $P_k$ in the main text, we assume that the traffic $T(j)$ of the $(k-1)$ existing centers is rougly the same: $T(j) \sim P / (k-1)$. A simple argument supports this assertion: in order for the $k^{th}$ center to appear, Eq. 7 in the main text must be satisfied, i.e. $Z_{ik}$ must be larger than all the $Z_{ij}\;j\in\left[1,k-1\right]$. Seen differently, it means that as long as a center $j$ is such that $Z_{ij}$ is larger than $Z_{ik}$, people will connect to that center. Therefore, all the centers $j$ are going to `fill up' until $Z_{ij}$ is too large. We can thus expect that, on average, all the centers will have the same traffic. Of course, the influence of space complicates this reasoning, and to convince the reader, we plot the distribution of $T(j)$ in the regime $\ell \gg L$ on Fig.~\ref{fig:traffic_distribution}.

We measure $\langle T(j) \rangle \approx 100 = P/N_c$ and the relative dispersion of the values is equal to $8\%$. Thus, our approximation of an equal repartition of people in the different centers in justified.

\subsubsection{Sublinear relation between $P_k$ and $k$} 

\paragraph{General argument}
The expression given in the main text for $P_k$ obviously relies on the fact that the $\eta_j$ are uniformly distributed, and it is legitimate to wonder how the scaling would be modified it their distribution was different. It is usefull to note that the previously derived exponent is the result of two contributions. First, $\left(k-1\right)$ from the assumption that the commuters are approximately equally distributed among the existing centers. We do not expect this to change with $\eta_j$'s distribution and thus we still have $T(j) \sim N/(k-1)$. Second, $\left(k-1\right)^{1/\mu}$ from the maximum of the difference $\eta_1-\eta_k$, which obviously depends on the distribution of $\eta$. Nevertheless, we can reasonably expect $\eta_1 -\eta_k = f(k,N_c)$ where $f$ is an increasing function of $k$ and decreasing function of $N_c$. If we only consider the evolution of $f$ with $k$, the function can be approximated around the current value of $k$ as:

$f \sim (k)^\theta$

with $\theta > 0$. It follows that we locally have:

\begin{equation}
k \sim P^{\frac{\mu}{\mu+\theta}}
\end{equation}

Which is still sublinear. The value of $\theta$ might change with $k$ depending on the distribution, but it will always be positive. Thus, the dependence of $k$ with $P$ might not be a power-law, but it will definitely be sublinear.

\paragraph{Illustration in the case of a Pareto distribution.}In order to illustrate this point, we compute the average of $\displaystyle\max_{j\in\left[1,k-1\right]} (\eta_j)-\eta_k = \eta_1 - \eta_k$, that is to say $\mathbb{E}\left[\eta_1 - \eta_k \right]$ for the Pareto distribution.

We assume that the random variables possess a probability density function $f(x) = x^{-1-\mu}$ defined on $\left[ 1, +\infty \right[$, and a cumulative distribution function $F(x) = 1 - x^{-\mu}$ defined on the same interval. The probability density function of the $r^{th}$ \emph{largest} sampled value $\eta_r$ is given by:

\begin{equation}
f_{\eta_r}(x) = \frac{N_c!}{(r-1)!(N_c-r)!} \left[ 1 - F(x) \right]^{r-1} F(x)^{N-r} x
\end{equation}

And $\mathbb{E}\left(\eta_r\right)$ is given by:

\begin{equation}
\mathbb{E}\left[ \eta_r \right] = \int_1^{+\infty} x\, f_{\eta_r}(x) \, dx
\end{equation} 

Which after calculations we find to be:

\begin{equation}
\mathbb{E}\left[ \eta_r \right] = \frac{\Gamma\left(N+1\right)\,\Gamma\left(r-\frac{1}{\mu}\right)}{\mu\, \Gamma\left(N+1-\frac{1}{\mu}\right)\, \Gamma\left(r\right)}
\end{equation}

Where we denote by $\Gamma$ the usual Gamma function. Using the Stirling approximation for the terms which have a dependance in $r$, we show that in the leading order:

\begin{equation}
\mathbb{E}\left[ \eta_r \right] = \frac{\Gamma\left(N+1\right)}{\mu\, \Gamma\left(N+1-\frac{1}{\mu}\right)} \; r^{-1/\mu}
\end{equation}

$\mathbb{E}\left[ \eta_r \right]$ is thus a decreasing function of $r$. Therefore, $\mathbb{E}\left[ \eta_1 - \eta_r \right]$ is an increasing function of $r$. Thus, assuming that the values of $\eta$ are drawn from a Pareto distribution, we still have a sublinear dependance of the number of subcenters in the population.

\subsubsection{Numerical verification of the expression of $P_k$}

We show in the main text that the value of the population $P_k$ at which the $k^{th}$ subcenter appears is given by:

\begin{equation}
\label{eq:argument_Nk}
P_k = P^* \left(k-1\right)^{\frac{\mu+1}{\mu}}
\end{equation}

\begin{figure}
\centering
\includegraphics[width=0.45\textwidth]{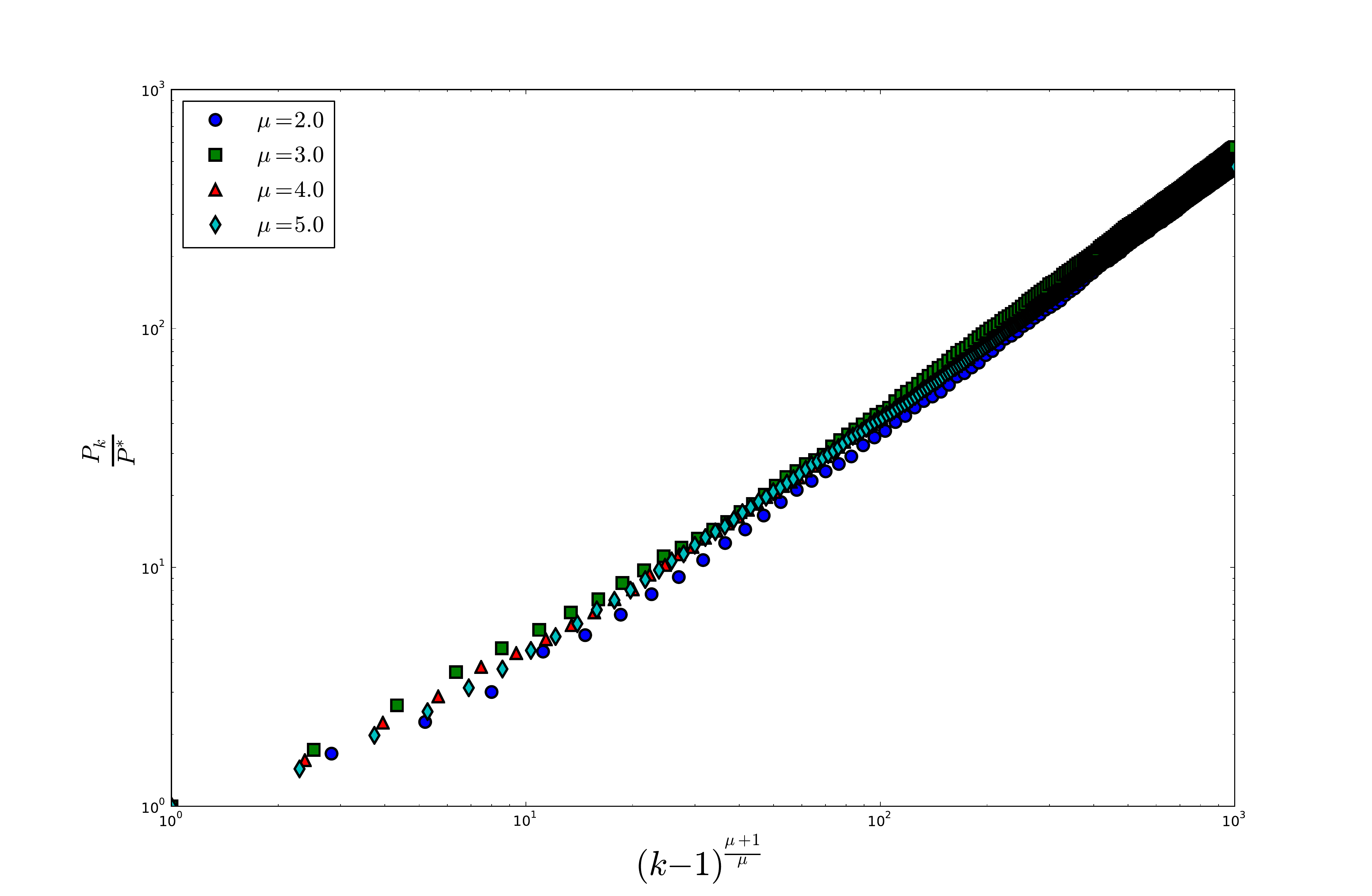}
\caption{Verification of the expression for $P_k$. Shows the average of the measured $P_k$ on many simulated configurations of the system as a function of $\left(k-1\right)^{\frac{\mu}{1+\mu}}$, for different values of the parameters. The points all collapse on the same curve, confirming our expression for $P_k$. The plots have been obtained for $N_c=10^3$, $L=1$, $c=1$ and $\ell=10^5$, in the attractivity-driven polycentric regime.~\label{fig:collapse_Nk}}
\end{figure}

In order to verify this formula, we numerically generate many configurations of the system and we measure the mean $P_k$ for several values of k and several values of $\mu$. We then plot the measured average $P_k/P^*$ as a function of $\left( k-1 \right)^{\frac{\mu+1}{\mu}}$ (Fig.~\ref{fig:collapse_Nk}) and see that all the plots collapse on the same curve, confirming the accuracy of our expression for $P_k$.

\subsection{Data analysis}

\subsubsection{Definition of the number of subcenter}

Extracting the number of activity centers from the Zip Code Business Patterns (ZBP) data is a priori a non-trivial task. ZBP data provide, among other things, the total number of employees per Zip Code Area in the United States every year between 1994 and 2010. Because Zip Code areas can have different areas, we first decide to normalise the number of employees to the area and obtain the employment density per Zip Code area.\\
If we further sort the employment densities in decreasing order and plot the employment densities of the different Zip Code areas as a function of their rank in the ordering (Supplementary Figure 3), we obtain a curves which decreases faster than exponentially for the first ranks, and then exponentially. We interpret this result as there existing a natural scale in the number of subcenters. Indeed, if the decrease is exponential, one can write for the employment density:

\begin{equation}
\rho_{e} \propto e^{-r/r_0}
\end{equation}

where $r_0$ is the typical value of the rank in this situation, and that we would interpret here as the number of subcenters. However, the rank-plot is not strictly exponential and we therefore define the number of subcenters using a threshold value $\alpha$. If we denote by $\rho_0$ the highest employee density in the city, we define $\rho_m$ as:

\begin{equation}
\rho_m = \frac{\rho_0}{\alpha}
\end{equation}

\begin{figure}
\centering
\includegraphics[width=0.45\textwidth]{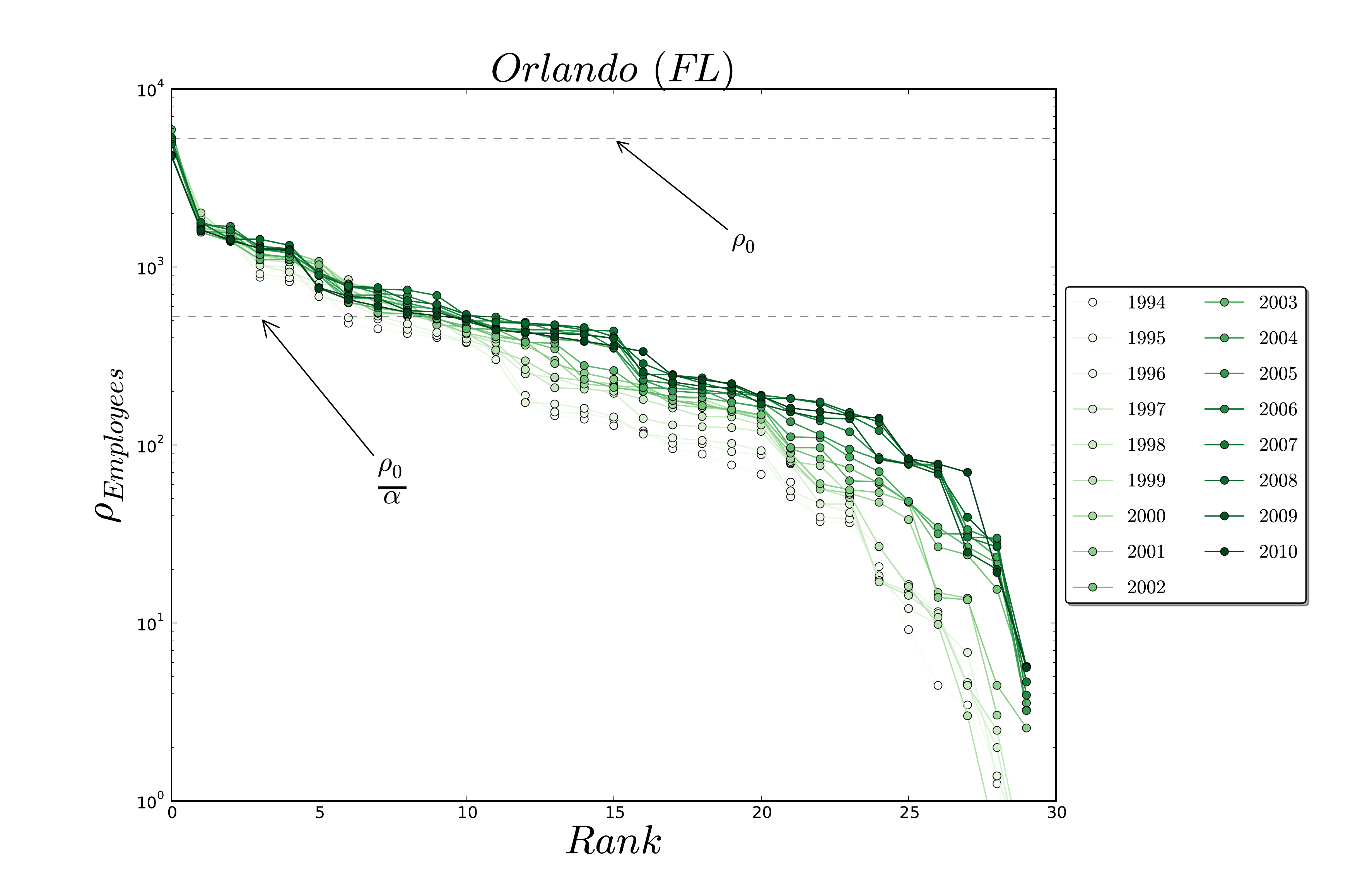}
\caption{Illustration of the definition of the number of subcenters. Shows the rank-plot of the employment density (in employees per square km) for Orlando(FL) between 1994 and 2010. We represent the largest employment density $\rho_0$ and the threshold value of employment $\rho_0 / \alpha$ for $1994$ with two dashed lines. The number of subcenters is taken to be the number of values of the employment density between those two values.\label{fig:definition}}
\end{figure}

And the number of subcenters $k$ is taken to be equal to the number of values of $\rho_{e}$ such that $\rho_{e} \in \left[ \rho_m, \rho_0 \right].$ (Fig.~\ref{fig:definition})\\

In the case where the rank-plot is exactly exponential, we thus have:

\begin{equation}
\alpha = \frac{\rho_0}{\rho_m} = e^{m/r_0}
\end{equation}

where $m$ is the number of activity centers. It follows:

\begin{equation}
m = r_0 \, \ln \left( \alpha \right)
\end{equation}

The order of magnitude of the number of centers is given by $r_0$ and relatively small variations of $\alpha$ therefore do not affect greatly the number of subcenters.

\subsubsection{Some details}

In order to plot the population as a function of the number of subcenters, we first extract the number of subcenters for every city in the dataset and for every year between 1994 and 2010, as indicated in the previous section.\\
We then apply the following treatment to the data:
\begin{itemize}
	\item If a given city corresponds to a single Zip Code, we only keep one point for that city: its population and $k=1$ center;
	\item We perform a Kolmogorov-Smirnov~\cite{Massey:1951} test between the rank-plots of the employment density in 1994 and 2010 for the remaining cities. If the probability that the two curves are different is greater than a certain threshold $p_{KS}$, we keep all the points for the city. Otherwise, we just keep a single point. 
\end{itemize}

At the end of this process, we obtain points that can be understood as coming from different realisations of a city. The following section gives the robustness of the results with regards to the value of the threshold $p_{KS}$ we choose.

\subsubsection{Robustness of the empirical results} 

\paragraph{Choice of $\alpha$} The empirical results presented in the main text will a priori depend on the value of the $\alpha$ we choose. We summarize the results obtained when fitting $k=f(\overline{P})$ assuming a power-law dependance and a multiplicative noise in Table~\ref{table:robustness}.
We see that for a reasonable range of values for $\alpha$ we always have a sublinear behaviour, and that given the estimated variance of the measured exponents, they are all compatible with each other.

\begin{figure}[!h]
\centering
\begin{tabular}{|c|c|c|c|c|c|}
\hline
$\alpha$ & 5 & 10 & 15 & 20 & 30\\ \hline
$\beta$ & 1.53 & 1.56 & 1.35 & 1.37 & 1.37\\
$\sigma_{\beta}$ & 0.27 & 0.15 & 0.14 & 0.12 & 0.11\\
$R^2$ & 0.74 & 0.87 & 0.81 & 0.84 & 0.85\\
\hline
\end{tabular}
\caption{Robustness of the empirical results. Shows the result for the fit for different values of $\alpha$ and $p_{KS}=0.50$.\label{table:robustness}}
\end{figure}

\paragraph{Choice of $p_{KS}$} It is also legitimate to wonder whether the empirical value we found for the exponent is affected by the value of the threshold $p_{KS}$. We summarize the results obtained with $\alpha=10$ in Table~\ref{table:pKS}. 
Again, we obtain a sublinear behaviour whatever the choice for the threshold $p_{KS}$ and the different values are compatible with each other.

\begin{figure}[!h]
\centering
\begin{tabular}{|c|c|c|c|c|c|c|}
\hline
$p_{KS}$ & 0.5 & 0.6 & 0.7 & 0.8 & 0.9 & 1\\ \hline
$\beta$ & 1.56 & 1.50 & 1.52 & 1.56 & 1.64 & 1.66\\
$\sigma_{\beta}$ & 0.19 & 0.18 & 0.18 & 0.17 & 0.17 & 0.15\\
$R^2$ & 0.87 & 0.81 & 0.83 & 0.85 & 0.86 & 0.89\\
\hline
\end{tabular}
\caption{Robustness of the empirical results. Shows the results for the fit for different values of $p_{KS}$ and $\alpha=10$. \label{table:pKS}}
\end{figure}

\paragraph{Comments}

The exponent derived from the plot on Fig. 3 in the main text is very sensitive to the first (k=1) and last point (k=26). Indeed, a fit excluding these points gives $\delta \sim 1.03 \pm 0.13\; (R^2 = 0.83)$. This exponent still agrees with the prediction of our model of $\beta < 1$, and would correspond to a very large value for $\mu \sim 32$ implying that the traffic dependence in Eq. 3 is a sharp threshold function. This indicates, if anything, that our empirical results should be confirmed through other studies based on different measures and datasets.

\bibliographystyle{prsty}

\end{document}